\documentclass[]{spie}
\usepackage{graphicx}

\makeatletter
\def\@tempa{revtex4}
\newif\ifisRevTeX
\ifx\class@name\@tempa
   \isRevTeXtrue
\else
   \isRevTeXfalse
\fi
\ifisRevTeX
   \renewcommand{\frontmatter@abstractheading}{%
      \begingroup\centering\large\vspace{2\baselineskip}%
      \abstractname\par\endgroup}
\fi
\makeatother

\usepackage{amssymb}

\newcommand{\EMB}{\boldsymbol}
\newcommand{\VEC}[1]{{\EMB{#1}}}
\newcommand{\LAB}[1]{{\mathsf{#1}}}
\newcommand{\FUN}[1]{{\mathrm{#1}}}
\newcommand{\MAT}[1]{{\mathbf{#1}}}
\newcommand{\TEN}[1]{{\mathfrak{#1}}}
\newcommand{\GRP}[1]{{\mathsf{#1}}}
\newcommand{\FLD}[1]{{\mathbb{#1}}}
\newcommand{\ALG}[1]{{\mathcal{#1}}}

\newcommand{\OL}{\overline}

\newcommand{\M}[1][1]{\hspace{#1em}}
\newcommand{\DAB}[1]{\rule[0pt]{0pt}{#1}}
\newcommand{\X}[1][1]{\DAB{#1ex}}

\newcommand{\HALF}{\mathchoice{{\textstyle\frac12}}{\frac12}{\frac12}{\frac12}}

\newcommand{\SIG}[1][{}]{\EMB\sigma_{\X[1.3]\M[-.05]#1}} % pain

\newcommand{\KET}[1]{|\,#1\,\rangle}
\newcommand{\BRA}[1]{\langle\,#1\,|}
\newcommand{\AVG}[1]{\langle\,#1\,\rangle}

\newtheorem{DEFINITION}{Definition}[section]

\usepackage[tbtags,nosumlimits]{amsmath}

\title{A Bloch-Sphere-Type Model for Two Qubits in the Geometric Algebra of a 6-D Euclidean Vector Space} 

\author{Timothy F.~Havel\supit{a} and Chris J.~L.~Doran\supit{b}
\skiplinehalf
\supit{a}Dept.~of Nuclear Engineering, Massachusetts Institute of Technology, \\
Cambridge, MA 02139, USA
\skiplinehalf
\supit{b}MRAO, Cavendish Laboratory, Cambridge University, Cambridge, \\
CB3 OHE, United Kingdom
}
\authorinfo{Further author information: T.F.H.: E-mail: tfhavel@MIT.EDU\\ 
C.J.L.D.: E-mail: C.Doran@MRAO.CAM.AC.UK}

\begin{document} \maketitle 

%%%%%%%%%%%%%%%%%%%%%%%%%%%%%%%%%%%%%%%%%%%%%%%%%%%%%%%%%%%%% 
\begin{abstract}
Geometric algebra is a mathematical structure that is inherent in any metric vector space, and defined by the requirement that the metric tensor is given by the scalar part of the product of vectors.
It provides a natural framework in which to represent the classical groups as subgroups of rotation groups, and similarly their Lie algebras.
In this article we show how the geometric algebra of a six-dimensional real Euclidean vector space naturally allows one to construct the special unitary group on a two-qubit (quantum bit) Hilbert space, in a fashion similar to that used in the well-established Bloch sphere model for a single qubit.
This is then used to illustrate the Cartan decompositions and subalgebras of the four-dimensional unitary group, which have recently been used by J.~Zhang, J.~Vala, S.~Sastry and K.~B.~Whaley [\textit{Phys.~Rev.~A} \textbf{67}, 042313, 2003] to study the entangling capabilities of two-qubit unitaries.
\end{abstract}

%>>>> Include a list of keywords after the abstract 

\keywords{Lie algebra, geometric algebra, quantum information}

%%%%%%%%%%%%%%%%%%%%%%%%%%%%%%%%%%%%%%%%%%%%%%%%%%%%%%%%%%%%%
\section{INTRODUCTION} \label{sect:intro}
The so-called Bloch sphere model of a two-level quantum system is widely used in quantum information processing, and as a visualization aid in quantum physics more generally.
Although its origins can be traced back to Riemann's stereographic projection of the unit sphere (minus one point) onto the complex plane, it has since been rediscovered several times in variety of physical contexts, including the Stokes vector in optics\cite{Stokes:1852}, the Bloch vector in nuclear magnetic resonance\cite{Bloch:46}, and the Feynman-Vernon-Hellwarth model of the maser\cite{FeyVerHel:57}.
This geometric interpretation of the state of a two-level quantum system is based on the fact that the special unitary group $\LAB{SU}(2)$ is the two-fold covering group of $\LAB{SO}(3)$, the proper rotations of a three-dimensional vector space, and the associated Lie algebra isomorphism $\LAB{so}(3) \approx \LAB{su}(2)$.
This isomorphism is ``accidental'' in the sense that it does not extend to an isomorphism of $\LAB{so}(n)$ with $\LAB{su}\big(\smash{\sqrt{1+n(n-1)/2}}\big)$ for some infinite sequence of integers $n > 3$.
It does, however, extend to the case $n=6$, i.e.~$\LAB{so}(6) \approx \LAB{su}(4)$, which corresponds to a pair of coupled two-level quantum systems or, from a quantum information processing perspective, a pair of \textit{qubits}.

\begin{figure}[bt]
\begin{center}
\includegraphics[width=6.5in]{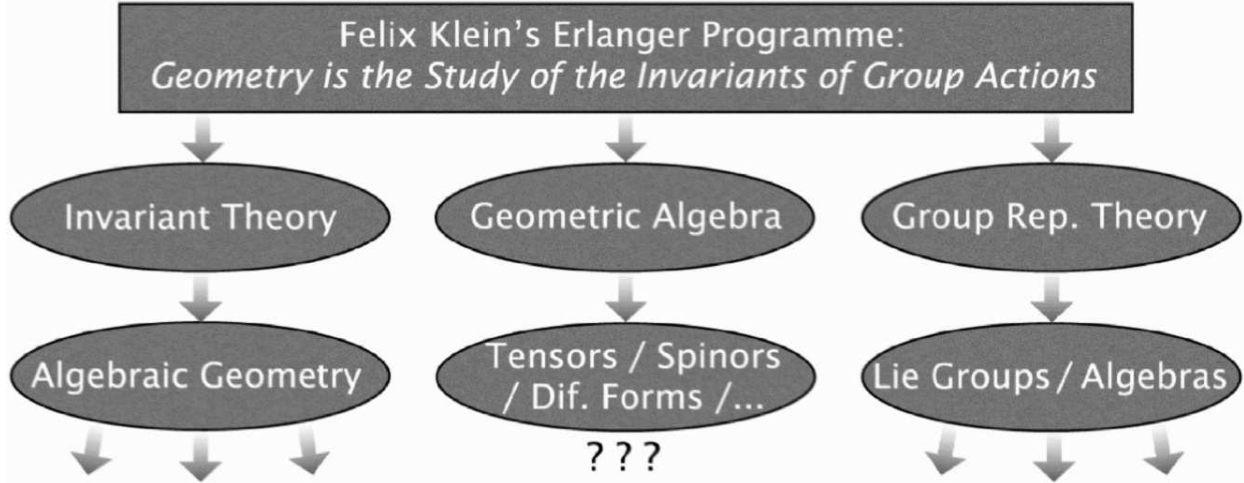}
\end{center}
\caption{ \label{fig:gahist}
The modern theory of geometry was initiated by Felix Klein's famous Erlanger programme of 1872, which promoted the view that ``geometry'' should be regarded as the study of the invariants of group actions.
This led to three main lines of mathematical research.
The first was known as invariant theory, and used polynomials in the coordinates to express geometric properties, and polynomials in these invariants (syzygies) to express geometric relations among the properties.
It proved too difficult to perform the necessary computations in these terms (at least before computers), and was basically subsumed into the more general theory of algebraic curves and surfaces now known as algebraic geometry.
The other was group representation theory, which was basically invented by Frobenius and was subsequently used to develop the abstract theory of Lie algebras developed by Lie and Engel.
Geometric algebra was actually initiated by Grassmann and, in the form of quaternions, by Hamilton even before Klein launched his programme, and subsequently developed by many well-known scientists and mathematicians including Clifford, Peano and Gibbs.
Due to a variety of historical accidents, it never became as well established as the other main formalisms, although its influence many be seen in several subsequent developments, e.g.~vector algebra, spinor theory and differential forms; in pure mathematics, it has more recently been used in the study Riemannian manifolds under the name of ``Clifford algebra''.\cite{LawsonMichel:89}
Over the last forty years or so, its diverse applications across physics and engineering have gradually become increasingly broadly recognized, starting with the work of Baylis, Hestenes and Sobczyk.\cite{Baylis:96,HesteSobcz:84,Hestenes:66,Hest!NF1:99}
A fascinating account of the history of these geometrical ideas may be found in a modern book by Isaac M.~Yaglom.\cite{Yaglom:88}
}
\end{figure}

In this article we construct a geometric model based on a six-dimensional Euclidean vector space for the states and transformations of a pair of qubits, and use this model to interpret a parametrization of $\LAB{SU}(4)$ that was recently derived from the basic theory of Lie algebras \cite{KhanejaGlaser:01,ZhVaSaWhal:03}.
It should first be noted that unlike the $\LAB{SU}(2)$ vs.~$\LAB{SO}(3)$ case, a six-dimensional vector space is not nearly large enough to fully parametrize the $15$-dimensional group $\LAB{SU}(4)$.
This would ordinarily be handled by introducing $6\times6$ rotation matrices parametrized by the six-dimensional analog of spherical-polar coordinates, but this does not illuminate the underlying geometry in any simple way.
Instead, therefore, we will be using a generalization of vector algebra to metric vector spaces of all dimensions and signatures, which is best known today as \emph{geometric algebra} \cite{Hestenes:03,DoranLasen:03}.
Although this is certainly much less widely known than matrix algebra, the intimate links between the structure of geometric algebra and the geometry of the underlying vector space makes it the ideal tool for our purposes.
Figure \ref{fig:gahist} attempts to put geometric algebra into historical perspective vis-\'a-vis the better known and related (though very different) formalisms of algebraic geometry and group representations.

This remaining sections of this article are organized as follows.
The first section below provides a brief introduction to geometric algebra, so as to make this article reasonably self-contained.
The next section shows how the geometric algebra of a three-dimensional Euclidean vector space applies to the Bloch sphere model.
This is extended to the six-dimensional model in the following section, using the fact that the even subalgebra of its geometric algebra is isomorphic to the usual algebra of complex $4\times4$ matrices.
Finally, we consider the parametrization of unitary transformations in the model that follows from the isomorphism $\GRP{so}(6) \approx \GRP{su}(4)$, along with possible extensions to more general completely positive linear maps.
In closing, we briefly discuss the connections of this work to other applications of geometric algebra to quantum mechanics, most notably the multiparticle space-time algebra.\cite{DorLasGul:93,SomLasDor:99,HavelDoran:02}

\section{BASIC CONCEPTS OF GEOMETRIC ALGEBRA} \label{sect:concepts}
Like most good mathematical ideas, geometric algebras may be defined in a number of different ways.
The following definition, taken from Ref.~\citenum{Lounesto:97}, has the advantage that it refers directly to operations and properties that we shall subsequently utilize.
\begin{DEFINITION}
An associative algebra over $\FLD R$ is the geometric algebra $\ALG G(p,q)$ of a metric vector space $\ALG V$ with nondegenerate quadratic form $Q: \ALG V \rightarrow \FLD R$ of signature $(p,q)$ if it contains both $\FLD R$ and $\ALG V$ as distinct subspaces such that:
\begin{itemize}
\item
the square of any vector $\VEC v \in \ALG V$ is $\VEC v^2 = Q(\VEC v)$;
\item
$\ALG V$ generates $\ALG G(p,q)$ as an algebra over $\FLD R$;
\item
$\ALG G(p,q)$ is not generated by any proper subspace of $\ALG V$.
\end{itemize}
\end{DEFINITION}
\noindent The algebra determined by these assumptions is unique up to isomorphism and, being semi-simple, also isomorphic to a direct sum of matrix algebras over a real division ring (i.e.~the reals, the complex numbers, or the quaternions).
Note further that in general the algebra $\ALG G(p,q)$ contains the vector space $\ALG V$ as a \emph{proper} linear subspace; the geometric interpretation of the entities in $\ALG G(p,q)$ that are not simply vectors in $\ALG V$ will be explained shortly.

As an immediate consequence of the first assumed property, the symmetric part of the \emph{geometric product} of two vectors is the value of the bilinear form defined by $Q$, since
\begin{equation}
\HALF \big( \VEC{ab} + \VEC{ba} \big) ~=~ \HALF \big( \VEC a^2 + \VEC b^2 - (\VEC a - \VEC b)^2 \big) ~=~ \HALF \big( Q(\VEC a) + Q(\VEC b) - Q(\VEC a - \VEC b) \big) .
\end{equation}
We shall henceforth refer to this as the \textit{inner product} of the vectors, which in keeping with modern conventions shall be denoted by $\VEC a \cdot \VEC b$.
The remaining, antisymmetric part of the geometric product, by way of contrast, is referred to as the \emph{outer product}, and written (following a notation that dates back all the way to Grassmann) as
\begin{equation}
\VEC a \wedge \VEC b ~\equiv~ \HALF \big( \VEC{ab} - \VEC{ba} \big) ~.
\end{equation}
This outer product may be extended to an \emph{associative} and totally antisymmetric product of any number of vectors via the recursive definition
\begin{equation}
\VEC a \wedge \VEC B_r ~\equiv~ \HALF \big( \VEC{aB}_r + {(-1)}^r \VEC B_r \VEC a \big) \M[3] \big( r = 1,2,\ldots \big) ,
\end{equation}
with $\VEC B_1 \equiv \VEC b \in \ALG V$.
These outer products span the entire real linear space on which $\ALG G(p,q)$ is built, and will be referred to in the following as \textit{$r$-blades}.

The geometric interpretation of an $r$-blade is both straightforward and natural.
It is the locus of the linear equation defined by the outer product of the blade with yet another vector, namely
\begin{equation}
L(\VEC B_r) ~\equiv~ \big\{ \VEC a \in \ALG V \mid \VEC a \wedge \VEC B_r ~=~ 0 \big\} ,
\end{equation}
and hence a \textit{subspace} of the underlying space of simple vectors $\ALG V$.
Indeed, the definition of the outer product immediately implies that $\VEC a \wedge \VEC b = 0$ if and only if $\VEC a = \alpha \VEC b$ for some scalar $\alpha$, and hence more generally that
\begin{equation}
\VEC a \wedge \VEC B_r ~=~ 0 \M\Leftrightarrow\M \VEC a ~=~ \alpha_1 \VEC b_1 +\cdots+ \alpha_r \VEC b_r \M[3] \big( \alpha_1,\ldots,\alpha_r \in \FLD R,~ \VEC B_r \equiv \VEC b_1 \wedge\cdots\wedge \VEC b_r \big) ,
\end{equation}
i.e.~$\VEC a$ is linearly dependent on the factors $\VEC b_1,\ldots,\VEC b_r \in \ALG V$ of $\VEC B_r$.
This subspace interpretation justifies the convention that the outer product of any $r > p + q$ vectors is identically zero.

A given orthonormal basis $\VEC e_{\X[1.4]1},\ldots,\VEC e_n$ of the underlying vector space $\ALG V$ ($n = p + q$) induces a basis for $\ALG G(p,q)$ as a real linear space, which is given by all combinations of outer products of basis vectors in lexicographic order, i.e.
\begin{equation}
\underset{1}{\underbrace{1_{\X}}}\,,~ \underset{n}{\underbrace{\VEC e_{\X[1.4]1},\ldots,\VEC e_{\X[1.4]n}}},~ \underset{\binom{n}{2}}{\underbrace{\VEC e_{\X[1.4]1} \wedge \VEC e_{\X[1.4]2} ,\ldots, \VEC e_{\X[1.4]n-1} \wedge \VEC e_{\X[1.4]n}}} ,~ \underset{\binom{n}{3}}{\underbrace{\VEC e_{\X[1.4]1} \wedge \VEC e_{\X[1.4]2} \wedge \VEC e_{\X[1.4]3} ,\ldots, \VEC e_{\X[1.4]n-2} \wedge \VEC e_{\X[1.4]n-1} \wedge \VEC e_{\X[1.4]n}}},~ \ldots,~ \underset{1}{\underbrace{\VEC e_{\X[1.4]1} \wedge\cdots\wedge \VEC e_{\X[1.4]n}}} ~.
\end{equation}
The numbers under each subsequence of $r$-blades is the dimension $\binom{n}{r}$ of the space of \textit{$r$-vectors} that they span, and thus the dimension of the entire algebra is
\begin{equation}
\sum_{r=0}^n \binom{n}{r} ~=~ 2^n ~.
\end{equation}
The algebra of outer products is in fact the well-known \textit{exterior algebra} $\bigwedge \ALG V$, which is widely used in the algebraic theory of projective geometry\cite{HodgePedoe:68}.
In keeping with this field, we will sometimes denote the subspace of $r$-vectors by $\bigwedge_r \ALG V$, and in order to avoid confusion with the dimension of $\ALG V$, the number of vectors $r$ in an $r$-blade will be called its \textit{grade}.

We now return to the inner product, and similarly extend it to all $r$-blades via the complementary definition
\begin{equation}
\VEC a \cdot \VEC B_r ~\equiv~ \HALF \big( \VEC{aB}_r - {(-1)}^r \VEC B_r \VEC a \big) \M[3] \big( r = 1,2,\ldots \big) .
\end{equation}
As an immediate consequence, we can express the geometric product of a vector and an $r$-blade as the sum of its inner and outer products,
\begin{equation}
\VEC{aB}_r ~=~ \VEC a \cdot \VEC B_r \,+\, \VEC a \wedge \VEC B_r ~.
\end{equation}
Unlike the outer product, however, the inner product is \emph{not} associative, and indeed it is readily verified that
\begin{equation}
\VEC a \cdot \big( \VEC a' \cdot \VEC B_r \big) ~=~ \big( \VEC a \wedge \VEC a' \big) \cdot \VEC B_r ~=~ - \big( \VEC a' \wedge \VEC a \big) \cdot \VEC B_r ~=~ - \VEC a' \cdot \big( \VEC a \cdot \VEC B_r \big) ~\ne~ \big( \VEC a \cdot \VEC a' \big) \cdot \VEC B_r ~.
\end{equation}
Indeed we will w.l.o.g.~set $\alpha \cdot \VEC B_r = 0$ for any scalar $\alpha$, since these products are not of much use otherwise.
Note in particular that $\VEC a \cdot (\VEC a \cdot \VEC B_r) = (\VEC a \wedge \VEC a) \cdot \VEC B_r = 0 \cdot \VEC B_r = 0$.

In addition to the inner and outer products, there are two unary involutions (mappings which applied twice give the identity) in geometric algebras.
The first of these is variously known as \textit{inversion} or the \emph{parity operation}, and denoted with an overbar, e.g.~$\OL{\!\VEC B}_r\,$.
This is defined on vectors as multiplication by $-1$, and extended by multilinearity to general $r$-blades, i.e.
\begin{equation}
\OL{\VEC b_1 \wedge\cdots\wedge \VEC b_r} ~\equiv~ {(-1)}^r\, \VEC b_1 \wedge\cdots\wedge \VEC b_r ~.
\end{equation}
It is easily verified that this operation respects the inner, outer and geometric products, i.e.~it is an algebra automorphism, so that the space of all inversion symmetric multivectors forms a subalgebra under the geometric product which is known as the \textit{even subalgebra} $\ALG G^+(p,q)$.
The other involution is known as \textit{reversion}.
This simply reverses the order of the vectors in every $r$-blade, and is denoted by a tilde, e.g.~$\tilde{\VEC B}_r\,$ or $\VEC B_r^\sim$.
Because the outer product is antisymmetric, the net effect is to change the sign of an $r$-vector whenever the parity of the permutation, as a product of transpositions, is odd, i.e.
\begin{equation}
\big( \VEC b_1 \wedge\cdots\wedge \VEC b_r \big)^\sim ~\equiv~ {(-1)}^{\binom{r}2}\, \VEC b_1 \wedge\cdots\wedge \VEC b_r ~.
\end{equation}
Reversion is not an automorphism but rather an anti-automorphism, meaning ${(\VEC{AB})}^\sim = \tilde{\VEC B}\tilde{\VEC A}$.

\begin{table}[tb] \begin{center}
\caption{ \label{tab:involutions} } \smallskip
\begin{tabular}{|l|cccccccc|} \hline
{} & \multicolumn8{|c|}{} \\[-2ex]
$r$ & 1 & 2 & 3 & 4 & 5 & 6 & 7 & 8 \\[1ex]
$r(r-1)/2$ & 0 & 1 & 3 & 6 & 10 & 15 & 21 & 28 \\[1ex] \hline
{} & \multicolumn8{|c|}{} \\[-2ex]
${(-1)}^r$ & $-$ & $+$ & $-$ & $+$ & $-$ & $+$ & $-$ & $+$ \\[1ex]
${(-1)}^{r(r-1)/2}$ & $+$ & $-$ & $-$ & $+$ & $+$ & $-$ & $-$ & $+$ \\[1ex] \hline
\end{tabular}
\end{center} \end{table}

Together, these two involutions allow one to say a good deal about the grade of an expression in the algebra.
For example, a quick look at their definitions shows that the inner and outer products of a vector with an $r$-vector change sign under inversion when $r$ is even but not when $r$ is odd so that, in particular, these products can have no $r$-vector part.
On the other hand, the inner product $\VEC a \cdot \VEC B_r$ changes sign under reversion when $r$ and $r(r-1)/2$ are both odd or both even but not otherwise, while for the outer product $\VEC a \wedge \VEC B_r$ it's the other way around.
The dependence of reversion and inversion symmetry on $r$ is summarized in Table \ref{tab:involutions}, which shows that these symmetries allow us to determine the value of $r\!\!\mod4$.
It further shows that the change in these symmetries on taking the inner and outer products of a vector with an $r$-vector is most simply explained by the fact that the former has rank $r-1$ and the latter $r+1$ (the latter was of course built into the definition of an $r$-blade).
Thus we say that the inner product is grade lowering and the outer grade raising.
In fact the inner and outer products of an $r$-blade with an $s$-blade are most simply defined as
\begin{equation}
\VEC A_r \cdot \VEC B_s ~\equiv~ \AVG{ \VEC A_r \VEC B_s }_{|r-s|} ~,\M[2]  \VEC A_r \wedge \VEC B_s ~\equiv~ \AVG{ \VEC A_r \VEC B_s }_{r+s} ~,
\end{equation}
where $\AVG{}_r$ denotes the orthogonal projection of the enclosed expression onto the subspace of $r$-vectors.

In much of what follows, we shall be using geometric algebra as a means of representing Lie algebras and groups.
The utility of geometric algebra in this regard stems from the fact that the commutator product of bivectors (aka $2$-vectors), which we shall write as
\begin{equation}
\VEC A_2 \times \VEC B_2 ~\equiv~ \VEC A_2 \VEC B_2 \,-\, \VEC B_2 \VEC A_2 ~,
\end{equation}
is readily shown by its symmetry with respect to inversion and reversion to yield another bivector.
In fact the bivector algebra under the commutator product is isomorphic to the Lie algebra of the special orthogonal group (isometries) of the underlying metric vector space, $\GRP{SO}(p,q)$.
The exponential map, likewise defined using the geometric product, yields a \textit{rotor} $R = \exp( \VEC B ) \in G^+(p,q)$ which rotates vectors by conjugation, i.e.~$\VEC a' = R\, \VEC a\, \tilde R$ (in the remainder of this paper we shall usually drop the grade subscript from our blades, indicating grade instead by $\VEC B = \AVG{\VEC B}_r\,$.
The Lie algebras of the other three main series in their classification, i.e.~the symplectic, unitary and complex orthogonal, can all be obtained as subalgebras of the orthogonal Lie algebras of higher dimensions.
In fact, the general linear group $\LAB{GL}(n)$ and its Lie algebra may be obtained by restricting the action of the isometries in $\ALG G(n,n)$ to the $n$-dimensional subspace of null vectors, and the classification theory for semisimple Lie algebras worked out within this framework.\cite{DoHeSoVA:93}

\section{QUBIT MECHANICS AND THE GEOMETRIC ALGEBRA OF 3-D SPACE}
This section is devoted to a fairly detailed exposition of the geometric algebra of a three-dimensional vector space, which will be denoted as $\ALG G(3) \equiv \ALG G(3,0)$.
This provides an excellent nontrivial example of most of the foregoing generalities on geometric algebras.
In addition, via the Bloch sphere model of a qubit's state as a three-dimensional unit vector, it will enable us to make a direct connection with quantum mechanics.

Let $\VEC e_{\X[1.4]1}, \VEC e_{\X[1.4]2}, \VEC e_{\X[1.4]3} \in \ALG V \approx \GRP R^3$ be an orthonormal basis for a three-dimensional Euclidean space, so that
\begin{equation}
\VEC e_{\X[1.4]1}^2 ~=~ \VEC e_{\X[1.4]2}^2 ~=~ \VEC e_{\X[1.4]3}^2 ~=~ 1 \M\text{and}\M \VEC e_i\, \VEC e_j ~=~ \VEC e_i \cdot \VEC e_j + \VEC e_i \wedge \VEC e_j ~=~ \VEC e_i \wedge \VEC e_j ~=\; -\VEC e_j \wedge \VEC e_i  ~=\; -{(\VEC e_i\, \VEC e_j)}^\sim
\end{equation}
for all $1 \le i < j \le 3$.
Using these relations, it is easily shown that the elements of the induced basis for the space of bivectors, denoted by
\begin{equation}
\VEC E_1 ~\equiv~ \VEC e_{\X[1.4]2}\, \VEC e_{\X[1.4]3} \,,~ \VEC E_2 ~\equiv~ \VEC e_{\X[1.4]3}\, \VEC e_{\X[1.4]1} \,,~ \VEC E_3 ~\equiv~ \VEC e_{\X[1.4]1}\, \VEC e_{\X[1.4]2} \,,
\end{equation}
likewise anticommute but square to $-1$, e.g.
\begin{equation}
\VEC E_1^2 ~=~ -{(\VEC e_{\X[1.4]2}\, \VEC e_{\X[1.4]3})}^\sim (\VEC e_{\X[1.4]2}\, \VEC e_{\X[1.4]3}) ~=~ -\VEC e_{\X[1.4]3} (\VEC e_{\X[1.4]2}^2) \VEC e_{\X[1.4]3} ~=~ -\VEC e_{\X[1.4]3}^2 ~=~ -1
\end{equation}
and
\begin{equation}
\VEC E_1 \VEC E_2 ~=~ (\VEC e_{\X[1.4]2}\M[0.05]  \VEC e_{\X[1.4]3}) (\VEC e_{\X[1.4]3}\M[0.05]  \VEC e_{\X[1.4]1}) ~=~ -\VEC e_{\X[1.4]1}\M[0.05] \VEC e_{\X[1.4]2} ~=~ -\VEC E_2 \VEC E_1 ~.
\end{equation}
Similarly, it can be shown that $\VEC E_1 \VEC E_2 \VEC E_3 = 1$.

These relations are, up to sign, exactly those which define the quaternion units $\VEC i$, $\VEC j$, $\VEC k$, and the even subalgebra $\ALG G^+(3)$ generated by the bivector  units is isomorphic to the full quaternion algebra. 
Indeed $\ALG G(3)$ constitutes a seamless merger of vector algebra with the quaternions.
The rotor (quaternion) that rotates vectors by an angle $\vartheta$ in the plane of $\VEC E_1$ is
\begin{multline}
R(\vartheta) ~\equiv~ \exp\!\big( (\vartheta/2)\, \VEC E_1 \big) ~=~ \sum_{k=0}^\infty \frac{(\vartheta/2)^k\, \VEC E_1^k}{k!} ~= \cdots \\
\sum_{k=0}^\infty \frac{(-1)^k (\vartheta/2)^{2k}}{(2k)!} \,+\, \VEC E_1 \sum_{k=0}^\infty \frac{(-1)^k (\vartheta/2)^{2k+1}}{(2k+1)!} ~=~ \cos(\vartheta/2) \,+\, \VEC E_1\, \sin(\vartheta/2) ~,~
\end{multline}
as is readily verified, e.g.
\begin{multline}
R(\vartheta)\, \VEC e_{\X[1.4]3}\, \tilde R(\vartheta) ~=~ \big( \cos(\vartheta/2) + \VEC E_1\, \sin(\vartheta/2) \big)\, \VEC e_{\X[1.4]3} \big( \cos(\vartheta/2) - \VEC E_1\, \sin(\vartheta/2) \big) ~=\cdots~ \\
\big( \cos(\vartheta/2) + \VEC E_1\, \sin(\vartheta/2) \big)^2\, \VEC e_{\X[1.4]3} ~=~ \big( \cos(\vartheta) + \VEC E_1\, \sin(\vartheta) \big)\, \VEC e_{\X[1.4]3} ~=~ \cos(\vartheta)\, \VEC e_{\X[1.4]3} + \sin(\vartheta)\, \VEC e_{\X[1.4]2} ~,~
\end{multline}
where we have used the relations $\VEC E_1\, \VEC e_{\X[1.4]3} = \VEC e_{\X[1.4]2} = -\VEC e_{\X[1.4]3}\, \VEC E_1$ (note that this is a left-hand rotation).
The appearance of the half-angle $\vartheta/2$ in this formula means that rotors constitute a spinor representation of $\LAB{SO}(3)$, which is a two-fold cover of $\LAB{SO}(3)$ since $\pm R(\vartheta)$ both yield the same rotation.

The ordered product of all three basis vectors $\VEC I \equiv \VEC e_{\X[1.4]1} \VEC e_{\X[1.4]2} \VEC e_{\X[1.4]3}$ is variously known as the unit trivector or \textit{pseudoscalar}.
This latter name stems from the fact that it (and all scalar multiples thereof) commute with all the basis vectors and hence with the entire algebra $\ALG G(3)$ they generate, since
\begin{equation}
\VEC I\, \VEC e_{\X[1.4]1} ~=~ \VEC e_{\X[1.4]1} \VEC e_{\X[1.4]2} \VEC e_{\X[1.4]3} \VEC e_{\X[1.4]1} ~=~ -\VEC e_{\X[1.4]1} \VEC e_{\X[1.4]2} \VEC e_{\X[1.4]1} \VEC e_{\X[1.4]3} ~=~ \VEC e_{\X[1.4]1}^2 \VEC e_{\X[1.4]2} \VEC e_{\X[1.4]3} ~=~ \VEC e_{\X[1.4]1}\, \VEC I ~,
\end{equation}
and similarly for $\VEC e_{\X[1.4]2}, \VEC e_{\X[1.4]3}\,$.
Furthermore, $\VEC I$ is yet-another square-root of $-1$, since
\begin{equation}
\VEC I^2 ~=~ -\VEC I \tilde{\VEC I} ~=~ -\VEC E_{\X[1.3]3}\, \VEC e_{\X[1.4]3}^2 \, \tilde{\VEC E}_{\X[1.3]3} ~=~ -\VEC E_3\, \tilde{\VEC E}_3 ~=~ \VEC E_3^2 ~=\; -1 ~.
\end{equation}
In other words, $\VEC I$ is algebraically identical to an abstract imaginary unit ``$\mathrm i$''.
The difference is that (like the basis bivectors above) the unit pseudoscalar admits a geometric interpretation as the \textit{oriented volume element} of space.
Its algebraic combination with real vectors, e.g.~$\VEC I\, \VEC e_{\X[1.4]1}$ as above, may also be interpreted as the oriented areal element in the plane orthogonal to $\VEC e_{\X[1.4]1}$.
This allows us to write the rotor for a right-handed rotation by an angle $\vartheta$ about an axis $\hat{\VEC a}$ ($\hat{\VEC a}^2 = 1$) a little more intuitively as
\begin{equation}
R_{\hat{\VEC a}}(\vartheta) ~=~ \exp\!\big( -\!\VEC I\, (\vartheta/2)\, \hat{\VEC a} \big) ~=~ \cos(\vartheta/2) \,-\, \VEC I\, \hat{\VEC a}\, \sin(\vartheta/2) ~.
\end{equation}
Finally, we have the relation
\begin{equation}
\VEC I\, \VEC E_\ell ~=~ -\VEC e_\ell \M[3] \big( 1 \le \ell \le 3 \big) ~,
\end{equation}
which implies that the cross product of vectors may be expressed in terms of their outer product as
\begin{equation}
\VEC a \times \VEC b ~=~ \tilde{\VEC I}\, ( \VEC a \wedge \VEC b ) ~.
\end{equation}

\begin{table}[b]
\vspace{-1ex}
\caption{ \label{tab:pauli} }
\begin{center}
\vspace{-1ex}
\begin{tabular}{|c|c|} \hline
{~} & {~} \\[-1.5ex]	% the things you have to do to get spaces right ...
the real numbers & real multiples of the $2\times2$ identity \\[1ex]
~three-dimensional space of vectors~ & traceless $2\times2$ Hermitian matrices \\[1ex]
~three-dimensional space of bivectors~ & traceless $2\times2$ anti-Hermitian matrices \\[1ex]
the pseudoscalars & imaginary multiples of the $2\times2$ identity \\[1ex]
\hline \end{tabular}
\end{center}
\vspace{-1ex}
\end{table}%

The state of a qubit is, of course, represented by a unit vector in a two-dimensional complex Hilbert space, which at first sight seems very far removed from anything in three-dimensional Euclidean geometry.
For someone who is already familiar with both, however, the connection may be stated in a single line: \emph{the Pauli matrix algebra is a representation of the geometric algebra $\ALG G(3)$.}
An isomorphism between them is obtained simply by mapping the unit vectors to the corresponding Pauli matrices, i.e.
\begin{equation}
\VEC e_{\X[1.4]1} ~\leftrightarrow~ \SIG[1] ~=~ \left[ \begin{smallmatrix} 0\,&\,1\\[1ex] 1\,&\,0 \end{smallmatrix} \right] ,\M \VEC e_{\X[1.4]2} ~\leftrightarrow~ \SIG[2] ~=~ \left[ \begin{smallmatrix} 0&-\mathrm i\\[1ex]\mathrm i&0 \end{smallmatrix} \right] ,\M \VEC e_{\X[1.4]3} ~\leftrightarrow~ \SIG[3] ~=~ \left[ \begin{smallmatrix} 1&0\\[1ex]0&-1 \end{smallmatrix} \right] ~,
\end{equation}
as may be verified by showing that these matrices satisfy the same algebraic relations as the basis vectors.
The relations between the multivectors in $\ALG G(3)$  and various types of $2\times2$ complex matrices are summarized in Table \ref{tab:pauli}.
In particular, the even subalgebra $\ALG G^+(3)$ (or quaternions if you prefer) is represented by all real multiples of $\LAB{SU}(2)$ matrices, i.e.~$\LAB{SU}(2) \approx \LAB{Spin}(3)$, the multiplicative group of rotors in $\ALG G^+(3)$.

As simple as this correspondence is, we have yet to say how one represents the state of a qubit in $\ALG G(3)$, and it is not obvious since they are usually represented by $2\times1$ vectors rather than $2\times2$ matrices.
The trick is to form a matrix with zeros in its second column, which has the same number of degrees of freedom in it as a $2\times1$ vector, and transforms correctly under left-multiplication by $\LAB{SU}(2)$, namely
\begin{equation}
\KET{\psi} ~=~ \begin{bmatrix} \psi_1 \\[1ex] \psi_2 \end{bmatrix} ~\leftrightarrow~ \begin{bmatrix} \;\psi_1 &  0~ \\[1ex] \;\psi_2 & 0~ \end{bmatrix} ~=~ \begin{bmatrix} \;\psi_1 & \!-\psi_2^* \\[1ex] \;\psi_2 & \psi_1^* \end{bmatrix} \M[-0.25] \begin{bmatrix} \;1 &  0\; \\[1ex] \;0 & 0\; \end{bmatrix} ~\leftrightarrow~ \Psi\, \HALF (1 + \VEC e_{\X[1.4]3}) ~=~ \Psi\, P_3 ~,
\end{equation}
\pagebreak[3]\par\noindent
where $\Psi \in \ALG G^+(3)$ and $P_3 \equiv \HALF (1 + \VEC e_{\X[1.4]3})$ is a projection operator, i.e.~$P_3^{\,2} = P_{\X[1.3]3\,}$.
This is described mathematically by saying that spinors form a \textit{left-ideal} in $\ALG G(3)$.

Having made this identification, we are now ready to complete the job of showing how qubit mechanics can be embedded within $\ALG G(3)$.
Expectation values are computed pretty much as usual, i.e.
\begin{equation}
\HALF\, \BRA\psi\, \MAT A\, \KET\psi ~\equiv~ \HALF\, \BRA\psi\,a_{\X[1.4]1} \SIG[1] + a_{\X[1.4]2} \SIG[2] + a_{\X[1.4]3} \SIG[3]\, \KET\psi ~=~ \AVG{P_3 \tilde{\Psi}\, \VEC a\, \Psi P_3}_{\X[1.4]0} ~=~ \AVG{\tilde{\Psi}\, \VEC a\, \Psi P_3}_{\X[1.4]0} ~,~
\end{equation}
where the vector $\VEC a \equiv a_1 \VEC e_{\X[1.4]1} + a_2 \VEC e_{\X[1.4]2} + a_3 \VEC e_{\X[1.4]3}\,$.
On the right-hand side of this equation, we have used $P_3^{\,2} = P_3$ together with the fact that the scalar part ``$\,\langle\,\rangle_{\X[1.4]0}$'' equals one-half the real part of the trace in the Pauli matrix representation.
This formula extends readily to ensembles of quantum systems in a mixed state if one defines the \emph{density operator} in $\ALG G(3)$ as
\begin{equation}
\rho ~\equiv~ 2\, \AVG{ \Psi P_3 \tilde{\Psi} } ~\implies~ \AVG{ \MAT A } ~=~ \AVG{ \rho\, \VEC a }_{\X[1.4]0} ~,
\end{equation}
where the unsubscripted angular brackets now refer to the ensemble average.
This latter formula admits a nice geometric interpretation if we recall the definition of $P_3$ and that $\Psi \in \ALG G^+(3)$ rotates vectors by conjugation, so that
\begin{equation}
\rho ~=~ 1 + \AVG{ \Psi \VEC e_{\X[1.4]3} \tilde\Psi } ~\equiv~ 1 + \VEC b ~\implies~ \AVG{ \MAT A } ~=~ \AVG{ \VEC a }_{\X[1.4]0} + \AVG{ \VEC{ba} }_{\X[1.4]0} ~=~ \AVG{ \VEC{ab} }_{\X[1.4]0} ~=~ \VEC a \cdot \VEC b ~.
\end{equation}
The vector $\VEC b$ of course corresponds to the usual Bloch vector, and represents the average state of the qubit over the ensemble.

This last equation implies that the rotor $\Psi$ can be interpreted as an instruction to rotate the reference vector $\VEC e_{\X[1.4]3}$ to the Bloch vector of the pure state in question.
This has rather interesting implications for the meaning of certain features of quantum mechanics.\cite{Hestenes:66,DoranLasen:03}
It shows that the indeterminate phase of the spinor is nothing but the angle of rotation about $\VEC e_{\X[1.4]3}\,$, explaining why it has no effect on the state.
It also shows why spinors transform with the half-angle of the corresponding Euclidean rotation quite simply.
The generalization of these interpretations to multi-qubit systems has received a considerable amount of attention over the last few years,\cite{DorLasGul:93,SomLasDor:99,HavelDoran:02} and we shall now revisit some of this work using a new geometric model for the states and operators of a two-qubit system.

\section{QUBIT MECHANICS AND THE GEOMETRIC ALGEBRA OF 6-D SPACE}
As mentioned in the Introduction, the Lie algebra isomorphisms $\LAB{spin}(6) \approx \LAB{so}(6) \approx \LAB{su}(4)$ imply that it should be possible to represent the states and operators of a two-qubit system within the geometric algebra of a six-dimensional Euclidean vector space $\ALG G(6)$.
We will now show by construction how this can actually be done, along with the geometric interpretations of the state of a pair of qubits which this isomorphism implies.

\subsection{The Lie and Matrix Algebra Isomorphisms}
The $15$-dimensional Lie algebra $\LAB{su}(4)$ is usually represented by $4\times4$ anti-Hermitian matrices, whereas we shall represent the elements of the Lie algebra $\LAB{spin}(6)$ by the $15$-dimensional subspace of bivectors in $\ALG G(6)$.
If we let $\SIG[k]^1 \equiv \SIG[k] \otimes \SIG[0]$, $\SIG[k]^2 \equiv \SIG[0] \otimes \SIG[k]$ be the tensor products of Pauli matrices with the $2\times2$ identity $\SIG[0]$ ($k = 1,2,3$), then it is readily verified that $\FUN i$ times these six matrices and their nine matrix products $\SIG[k]^1\SIG[\ell]^2 = \SIG[k] \otimes \SIG[\ell]$ span the $15$-dimensional space of anti-Hermitian matrices and so generate all of $\LAB{su}(4)$.
We shall also let $\VEC e_{\X[1.4]1}, \VEC e_{\X[1.4]2}, \VEC e_{\X[1.4]3}, \VEC f_{\!1}, \VEC f_{\!2}, \VEC f_{\!3}$ be an orthonormal basis of  $ \ALG V \approx \FLD R^6$, and abbreviate the bivectors of $\wedge_2 \ALG V$ by:
\begin{alignat}{2}
\VEC G_{\X[1.4]10} ~\equiv~ & \VEC e_{\X[1.4]2} \VEC e_{\X[1.4]3} &\M[2]& \text{(and all cyclic permutations over the indices $1,2,3$)} \nonumber \\
\VEC G_{\X[1.4]01} ~\equiv~ & \VEC f_{\!2} \VEC f_{\!3} &\M[2]& \text{(and all cyclic permutations over the indices $1,2,3$)} \\ \nonumber
\VEC G_{ij} ~\equiv~ &\VEC e_{\X[1.4]\M[0.05]i\M[0.05]} \VEC f_{\!j}\, {(-1)}^{\smash{\delta_{ij}}\,} &\M[2]& \text{(for all $1 \le i, j \le 3$, where $\delta$ is the Kronecker delta).}
\end{alignat}
Then it is easily verified that under the correspondence indicated below the matrix and basis bivectors satisfy the same commutation relations, so that the linear mapping defined by this correspondence among the basis elements extends to real linear mapping between these two spaces that constitutes an instance of the purported isomorphism:
\begin{alignat}{3}
\FUN i\, \SIG[1]^1~ & \leftrightarrow~ \VEC G_{\X[1.5]10} &\M[4] \FUN i\, \SIG[2]^1~ & \leftrightarrow~ \VEC G_{\X[1.5]20} &\M[4] \FUN i\, \SIG[3]^1~ & \leftrightarrow~ \VEC G_{\X[1.5]30} \nonumber \\[0.5ex]
\FUN i\, \SIG[1]^2~ & \leftrightarrow~ \VEC G_{\X[1.5]01} &\M[4] \FUN i\, \SIG[2]^2~ & \leftrightarrow~ \VEC G_{\X[1.5]02} &\M[4] \FUN i\, \SIG[3]^2~ & \leftrightarrow~ \VEC G_{\X[1.5]03} \nonumber \\[0.5ex]
\FUN i\, \SIG[1]^1\SIG[1]^2~ & \leftrightarrow~ \VEC G_{\X[1.5]11} & \FUN i\, \SIG[2]^1\SIG[1]^2~ & \leftrightarrow~ \VEC G_{\X[1.5]21} & \FUN i\, \SIG[3]^1\SIG[1]^2~ & \leftrightarrow~ \VEC G_{\X[1.5]31} \\[0.5ex] \nonumber
\FUN i\, \SIG[1]^1\SIG[2]^2~ & \leftrightarrow~ \VEC G_{\X[1.5]12} & \FUN i\, \SIG[2]^1\SIG[2]^2~ & \leftrightarrow~ \VEC G_{\X[1.5]22} & \FUN i\, \SIG[3]^1\SIG[2]^2~ & \leftrightarrow~ \VEC G_{\X[1.5]32} \\[0.5ex] \nonumber
\FUN i\, \SIG[1]^1\SIG[3]^2~ & \leftrightarrow~ \VEC G_{\X[1.5]13} & \FUN i\, \SIG[2]^1\SIG[3]^2~ & \leftrightarrow~ \VEC G_{\X[1.5]23} & \FUN i\, \SIG[3]^1\SIG[3]^2~ & \leftrightarrow~ \VEC G_{\X[1.5]33}
\end{alignat}
Specifically, the bivector basis elements satisfy the following commutation relations:
\begin{alignat}{2}
\VEC G_{\X[1.4]ik} \times \VEC G_{\X[1.4]ik} & ~=~ 0 &\M[4]& \text{(for all $0 \le i,k \le 3$)} \nonumber \\
\VEC G_{\X[1.4]ik} \times \VEC G_{j\ell} & ~=~ 0 &\M[4]& \text{(for all $1 \le i,j,k,\ell \le 3$ with $i \ne j$ and $k \ne \ell$)} \nonumber \\
\VEC G_{\X[1.4]1k} \times \VEC G_{\X[1.4]2k} & ~=~ \!-\VEC G_{\X[1.4]30} &\M[4]& \text{(for all $0 \le k \le 3$ and cyclic permutations over the first index)} \\ \nonumber 
\VEC G_{\X[1.4]i1} \times \VEC G_{\X[1.4]i2} & ~=~ \!-\VEC G_{\X[1.4]03} &\M[4]& \text{(for all $0 \le i \le 3$ and cyclic permutations over the second index)} \\ \nonumber
\VEC G_{\X[1.4]10} \times \VEC G_{\X[1.4]2\ell} & ~=~ \!-\VEC G_{\X[1.4]3\ell} &\M[4]& \text{(for all $1 \le \ell \le 3$ and cyclic permutations over the first index)} \\ \nonumber
\VEC G_{\X[1.4]01} \times \VEC G_{\X[1.4]j2} & ~=~ \!-\VEC G_{\X[1.4]j3} &\M[4]& \text{(for all $1 \le j \le 3$ and cyclic permutations over the second index)}
\end{alignat}

A little further inspection quickly shows that not only do these matrices and bivectors correspond under the commutator product, but within any three-dimensional subspace the products of these matrices coincides with the geometric product of the bivectors.
Since the reversion of bivectors corresponds to Hermitian transposition in the matrix algebra, reversion symmetric entities in the \emph{even} subalgebra $\EMB{\ALG G}^+(6)$ serve as the analogs of Hermitian matrices in our six-dimensional model.
These entities consist of the scalars and the $4$-vectors, which together span a real $16$-dimensional subspace as they should.
The final dimension in the even subalgebra is the pseudoscalar $\VEC I \equiv \VEC e_{\X[1.4]1} \VEC e_{\X[1.4]2} \VEC e_{\X[1.4]3} \VEC f_{\!1} \VEC f_{\!2} \VEC f_{\!3}$.
This squares to $-1$, commutes with everything in $\EMB{\ALG G}^+(6)$ and maps $2$-vectors to $4$-vectors by multiplication (and vice versa).
In short, it is clearly a geometric analog of the imaginary unit times the identity in the $4\times4$ matrix algebra.

The next task is to find an analog of the state vectors of pure states, which we shall again regard as a left-ideal, i.e.~as matrices with all their entries equal to zero outside their first column, in analogy with the state vectors of single qubits in $\EMB{\ALG G}(3)$ (see above).
Since the tensor products of single qubit pure states span the full two-qubit Hilbert space in the matrix model, it is reasonable to use the analogous construction here, i.e.
\begin{equation}
\KET{\phi} \otimes \KET{\varphi} ~\leftrightarrow~ \Phi^1 P_3^1\, \Theta^2 P_3^2 ~\equiv~ \Phi^1 \HALF \big( 1 - \VEC I \VEC e_{\X[1.4]1} \VEC e_{\X[1.4]2} \big) \Theta^2 \HALF \big( 1 - \VEC I \VEC f_{\!1} \VEC f_{\!2\,} \big) ~=~ \Phi^1 \Theta^2 \tfrac14 \big( 1 - \VEC I\VEC G_{30} - \VEC I\VEC G_{03} - \VEC I\VEC G_{33} \big) ,
\end{equation}
where we have written the $4$-vector basis elements $\SIG[3]^1 \leftrightarrow \VEC e_{\X[1.4]3} 
\VEC f_{\!1} \VEC f_{\!2} \VEC f_{\!3\,}$ and $\SIG[3]^2 \leftrightarrow \VEC e_{\X[1.4]1} \VEC e_{\X[1.4]2} \VEC e_{\X[1.4]3} \VEC f_{\!3\,}$ as $-\VEC I$ times the complementary bivectors.
The multivectors $\Phi^1$ \& $\Theta^2 \in \EMB{\ALG G}^+(6)$ are actually the sums of scalars and bivectors in the even subalgebras $\EMB{\ALG G}^+(3)$ generated by $\VEC e_{\X[1.4]1}, \VEC e_{\X[1.4]2}, \VEC e_{\X[1.4]3}$ \& $\VEC f_{\!1}, \VEC f_{\!2}, \VEC f_{\!3}$, respectively.
Under the geometric product, these two four-dimensional even subalgebras generate the full $16$-dimensional even subalgebra, and the normalization condition $\Psi\tilde{\Psi} = 1$ on the state vectors $\Psi P_3^1 P_3^2 \in \EMB{\ALG G}^+(6)$ ensures that $\Psi$ corresponds to a unitary in the matrix algebra.
In terms of these matrices, each state vector is identified with the set of all unitary matrices with the same first column.

Geometric analogs of density matrices, henceforth called density operators to distinguish them, may be constructed in exactly the same fashion as in the matrix algebra as well.
The dyadic (sometimes misnamed the outer!) product of a state vector with its Hermitian conjugate clearly corresponds to
\begin{equation}
\KET\psi\BRA\psi ~\leftrightarrow~ \Psi\, P_3^1 P_3^2\, \tilde{\Psi} ~.
\end{equation}
Just as in the Bloch sphere model of a single qubit, this may be regarded as a unitary transformation of the density operator of the reference state $P_3^1 P_3^2 \leftrightarrow \KET{00}\BRA{00} $ to that corresponding to $\KET\psi\BRA\psi$.
General density operators $\rho \in \EMB{\ALG G}^+(6)$ may then be obtained by taking convex combinations of those of pure states in the usual way.
A geometric characterization of the positive semidefiniteness of density matrices does not, alas, fall out immediately.
The best one can say is that powers of the density operator have scalar parts (four times the trace in the matrix representation) that satisfy certain inequalities. Letting $\hat\rho \equiv \rho - 1/4$ be the ``traceless part'' of $\rho$, the first of these inequalities are
\begin{equation}
\tfrac3{16} ~\ge~ \AVG{ \hat\rho^2 }_{\X[1.5]0} ~,\M \tfrac3{32} ~\ge~ \AVG{ \hat\rho^3 }_{\X[1.5]0} ~,\M \tfrac{15}{128} ~\ge~ \AVG{ \hat\rho^4 }_{\X[1.5]0} ~. 
\end{equation}

Somewhat more insight into the geometry of pure states can be obtained from the well-known Schmidt decomposition.
This is usually written as a convex combination of tensor products of one-qubit states, but with a little work one can parametrize all the quantities appearing therein in terms of angles plus an overall scale factor, i.e.
\begin{multline}
\KET\psi ~=~ \varrho\, e^{-\FUN i\phi}\, \Bigg( \cos(\varsigma/2)\, e^{-\FUN i \tau / 2} \begin{bmatrix} \cos(\vartheta^{1\!}/2) e^{-\FUN i\varphi^{1\!}/2} \\[1ex] \sin(\vartheta^{1\!}/2) e^{\, \FUN i\varphi^{1\!}/2} \end{bmatrix} \otimes \begin{bmatrix} \cos(\vartheta^{2\!}/2) e^{-\FUN i\varphi^{2\!}/2} \\[1ex] \sin(\vartheta^{2\!}/2) e^{\, \FUN i\varphi^{2\!}/2} \end{bmatrix} ~+ \cdots \\
\sin(\varsigma/2)\, e^{\,\FUN i \tau / 2} \begin{bmatrix} \sin(\vartheta^{1\!}/2) e^{-\FUN i\varphi^{1\!}/2} \\[1ex] -\cos(\vartheta^{1\!}/2) e^{\, \FUN i\varphi^{1\!}/2} \end{bmatrix} \otimes \begin{bmatrix} \sin(\vartheta^{2\!}/2) e^{-\FUN i\varphi^{2\!}/2} \\[1ex] -\cos(\vartheta^{2\!}/2) e^{\, \FUN i\varphi^{2\!}/2} \end{bmatrix} \Bigg) .
\end{multline}
The analog of this decomposition in $\EMB{\ALG G}^+(6)$ is
\begin{equation}
\Psi ~=~ \varrho\, e^{-\VEC I\phi}\, e^{-\varphi^{1\!}/2\, \VEC G_{30} - \varphi^{2\!}/2\, \VEC G_{03}}\, e^{-\vartheta^{1\!}/2\, \VEC G_{20} - \vartheta^{2\!}/2\, \VEC G_{02}}\, e^{-\tau/2\, (\VEC G_{30}  + \VEC G_{03})}\, e^{-\varsigma/2\, \VEC G_{22}}\, P_3^1 P_3^2 ~.
\end{equation}
Although this may at first glance appear formidable, the interpretation of each factor in it is actually fairly straightforward.
In left-to-right order, first two parameters are the overall magnitude $\varrho$ (usually set to unity) and overall phase angle $\phi$.
The next factor contains the azimuthal angles $\varphi^1, \varphi^2$ of the Bloch vectors of the two qubits, as rotations in the planes of the bivectors $\VEC G_{30}, \VEC G_{03}$ respectively.
The next two contain the altitudinal angles $\vartheta^1, \vartheta^2$ of the Bloch vectors as rotations in the $\VEC G_{20}, \VEC G_{02}$ planes.
Thereafter we have the parameter $\tau$ which equals the \emph{sum} of the two angles of rotation about the reference directions of the two Bloch vectors, because the difference $\VEC G_{30} - \VEC G_{03}$ is projected to zero by the idempotents $P_3^1 P_3^2$.
The corresponding exponential rotates the two components obtained on expansion of the last factor in opposite directions, namely
\begin{equation}
\exp\!\big(\! -\varsigma/2\, \VEC G_{22} \big) ~=~ \cos(\varsigma/2) \,-\, \VEC G_{22}\, \sin(\varsigma/2) ~.
\end{equation}
The entanglement of the state is usually measured in terms of the entropy
\begin{equation}
-\cos(\varsigma/2) \log\!\big( \cos(\varsigma/2) \big) - \sin(\varsigma/2) \log\!\big( \sin(\varsigma/2) \big) ~,
\end{equation}
and depends only upon this last angle.

As an example, we consider the maximally entangled singlet state $\KET\psi \equiv (\KET{10} - \KET{01})/\sqrt2$, which has the Schmidt parameters
\begin{equation}
\varrho ~=~ 1 \,,~ \phi ~=~  \tau ~=~  \varphi^1 ~=~ \varphi^2 ~=~ \vartheta^1 ~=~ 0 \,,~ \vartheta^2 ~=~ \pi \,,~ \varsigma ~=~ -\pi/2 ~.
\end{equation}
It follows that the state is represented by
\begin{equation}
\Psi ~\equiv~ -\VEC G_{02}\, \tfrac1{\sqrt2} \big( 1 + \VEC G_{22} \big)\, P_3^1 P_3^2 ~=~ \tfrac1{\sqrt2} \big( \VEC G_{20} - \VEC G_{02} \big) ~,
\end{equation}
which corresponds to the density operator
\begin{equation} \label{eq:singlet}
\Psi\, P_3^1 P_3^2\, \tilde\Psi ~=~ \tfrac14 \big( 1 - \VEC I \VEC G_{11} - \VEC I \VEC G_{22} - \VEC I \VEC G_{33} \big)
\end{equation}
The next section will show that this is invariant under simultaneous rotations of both qubits together.

\begin{figure}[tb]
\begin{center}
\includegraphics[width=3.5in]{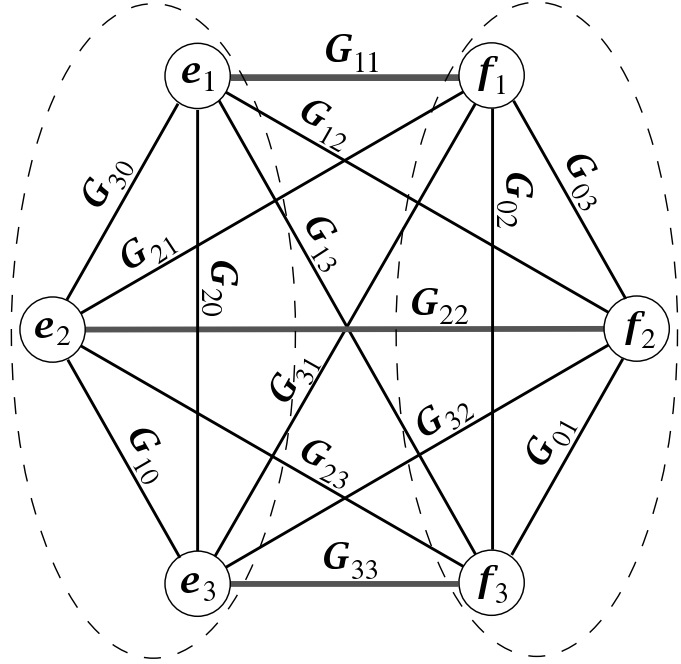}
\end{center}
\caption{ \label{fig:2qubits6d}
Combinatorial graph in which the vertices correspond to the basis vectors for a Bloch sphere model of each of the two qubits ($[\VEC e_{\X[1.4]1}, \VEC e_{\X[1.4]2}, \VEC e_{\X[1.4]3}]$ \& $[\VEC f_{\!1}, \VEC f_{\!2}, \VEC f_{\!3}]$, respectively) and the edges to the corresponding bivector basis $\VEC G_{ij\,}$.
The dashed ellipses enclose the induced subgraphs which correspond to the ``local'' subalgebras of the two Bloch sphere models, while the perfect matching of a Cartan subalgebra is indicated by the heavier lines on edges of $\VEC G_{11}, \VEC G_{22}, \VEC G_{33}$.
}
\end{figure}

\subsection{Cartan Decompositions and Subalgebras}
Thus far, we have not done anything with our six-dimensional model and geometric algebra that could not be done in the matrix algebra -- which is probably a good thing since the matrix algebra is evidently able to handle all of nonrelativistic qubit mechanics!
Nevertheless, we hope to show that the additional degrees of freedom present in our model do allow for some efficiency in notation, or at least serve a pneumonic in keeping track of commutation relations.
To this end we shall identity the basis vectors $\VEC e_{\X[1.4]1},\ldots,\VEC f_{\!3}$ with the vertices of a combinatorial graph, and the corresponding bivector basis $\VEC G_{ij}$ with the edges (pairs of vertices) in the graph, as depicted in Fig.~\ref{fig:2qubits6d}.
Any two nonidentical edges meeting at a common vertex then correspond to anticommuting bivectors, while any pair of edges with no vertex in common correspond to commuting bivectors.
Furthermore the commutator of any pair of bivectors, the edges of which have a vertex in common, equals the bivector of the edge connecting the other pair of vertices up to sign.
From this one sees immediately that the edges in any induced subgraph (i.e.~a subset of the vertices together with all the edges between them), as bivectors, generate a subalgebra $\TEN e$ of the full bivector algebra.
In addition, all the bivectors in this subalgebra commute with all those bivectors with their edges in the complementary induced subgraph $\TEN f$ (i.e.~that induced by the complement of the subgraph's vertex set),
and hence the product of these two subalgebras $\TEN g \equiv \TEN e \times \TEN f = \TEN f \times \TEN e = \TEN e \oplus \TEN f$ is again a subalgebra.
And finally, the commutator of any bivector in $\TEN g$ with a basis bivector, the edge of which connects a vertex of the subgraph to a vertex outside of it, is either zero or else a bivector outside $\TEN g$ of the same kind, while the commutator of any two basis bivectors outside $\TEN g$ is either zero or a bivector in the subalgebra.

Subalgebras of a Lie algebra which ``absorb'' their orthogonal complements like this are known as \textit{Cartan decompositions}.\cite{FuchsSchwe:97}
Their importance lies in the role they play in parametrizing Lie group of the full subalgebra, at least in some neighborhood of the identity.
To describe this fully, we need to also introduce the related concept of a \textit{Cartan subalgebra}, which is a subalgebra $\TEN h$ generated by a maximal set of commuting basis bivectors in $\TEN g^\perp$ (note that any noncommuting set of basis bivectors would not generate a subalgebra $\TEN h \subset \TEN g^\perp$).
Then if $\TEN l$ is the full Lie algebra in question and $\TEN G$, $\TEN H$ and $\TEN L$ are the Lie groups corresponding to the Lie algebras $\TEN g$, $\TEN h$ and $\TEN l$, respectively, we have
\begin{equation}
\TEN L ~=~ \TEN G\, \TEN H\, \TEN G ~,  \label{eq:cartan}
\end{equation}
where the juxtaposition of Lie groups on the right-hand side indicates the composition of all possible pairs of elements from these subgroups of $\TEN L$ in the given order.
For want of a better name, we shall call this a \textit{Cartan factorization} of the group.

The applications of Cartan factorizations to quantum control were first developed by Khaneja \& Glaser\cite{KhanejaGlaser:01}, and their implications for the geometry of nonlocal two-qubit operations has been nicely worked out by Zhang \textit{et al.}\cite{ZhVaSaWhal:03}, to which references the reader is directed for a full account.
Here we are mainly interested in what new insights our six-dimensional model and the graph theory notation for its bivectors can provide.

The most useful Cartan factorization, with regard to the ``local'' structure of $\GRP{SU}(4)$, is that indicated by the dashed ellipses in Fig.~\ref{fig:2qubits6d} as well as by our chosen notation.
Each of the two sets of basis vectors $\{\VEC e_{\X[1.4]i}\}$, $\{\VEC f_{\!j}\}$ generates a subalgebra $\ALG G(3)$ of $\ALG G(6)$, showing how the Bloch vector model of each qubit is embedded as a pair of complementary three-dimensional orthogonal subspaces in six-dimensions.
Any \textit{complete bipartite matching} between the corresponding complementary sets of three vertices each in the graph, i.e.~any set of three edges between the two sets with no vertices in common, then corresponds to a Cartan subalgebra; the most symmetric choice is $\VEC G_{11}, \VEC G_{22}, \VEC G_{33}$, as indicated in the figure.
Equation (\ref{eq:cartan}) then says that any rotation in six dimensions can be uniquely written as pair of rotations in complementary three-dimensional subspaces, rotations in three mutually orthogonal planes each intersecting both three-dimensional subspaces in a line, and another pair of rotations on the three-dimensional subspaces.
This is of course perfectly analogous to the parametrization of $\GRP{spin}(3)$ by Euler angles, in which a three-dimensional rotation is expressed as a rotation about the $\LAB z$-axis, followed by a rotation about the $\LAB x$-axis and finally a second $\LAB z$-rotation -- and which can also be derived from a Cartan decomposition.

As an example, consider the Cartan factorization of a unitary that maps between the local or computational basis of a four-dimensional Hilbert space and the corresponding basis of Bell states.
The matrix of this unitary versus the local basis has the Bell states as its columns, 
\begin{equation}
\MAT Q ~\equiv~ \frac1{\sqrt2} \begin{bmatrix} ~1&\M[0.25]0&0&\FUN i~ \\[1ex] ~0&\M[0.25]\FUN i&1&0~ \\[1ex] ~0&\M[0.25]\FUN i&\!-1&0~ \\[1ex] ~1&\M[0.25]0&0&\!-\FUN i\M[0.25] \end{bmatrix} ,
\end{equation}
and its Cartan factorization, up to an overall phase (exercise!), may be written as
\begin{multline}
\exp\!\big(\! -\vartheta (\MAT G_{01} - \MAT G_{02} + \MAT G_{03}) \big)\, \exp\!\big(\, \vartheta (\MAT G_{10} - \MAT G_{20} + \MAT G_{30}) \big)\,  \exp\!\big(\, (\pi/4)\, \MAT G_{33} \big)\, \cdots \\
\exp\!\big(\, (\pi/\sqrt8) (\MAT G_{20} - \MAT G_{30}) \big)\, \exp\!\big(\! - \vartheta (\MAT G_{01} + \MAT G_{02} + \MAT G_{03}) \big) , \M \label{eq:cartanQ}
\end{multline}
where $\vartheta = 4 \pi / \sqrt{27}$.
As noted in Ref.~\citenum{Makhlin:02}, this unitary transformation has the curious property of taking the Pauli matrices into $\FUN i$ times a set of anti-symmetric real matrices, which in turn generate the Lie algebra $\GRP{so}(4) \approx \GRP{so}(3) \oplus \GRP{so}(3)$.

Although it makes this isomorphism manifest in the standard Pauli matrix representation, $\MAT Q$ defines a relatively complicated mapping into the Bell basis.
Indeed, it seems there is no way to identify the subalgebra $\TEN g = \GRP{so}(3) \oplus \GRP{so}(3)$ with a four-dimensional subspace within our six-dimensional model.
A much simpler mapping into the Bell basis is defined by
\begin{equation}
\MAT Q' ~\equiv~ \frac1{\sqrt2} \begin{bmatrix} ~1&0&0&\M[0.25]1~ \\[1ex] ~0&1&1&\M[0.25]0~ \\[1ex] ~0&\!-1&1&\M[0.25]0~ \\[1ex] -1&0&0&\M[0.25]1~ \end{bmatrix} ,
\end{equation}
which has the Cartan factorization
\begin{multline}
\exp\!\big(\!-\vartheta (\MAT G_{01} + \MAT G_{02} + \MAT G_{03}) \big)\, \exp\!\big(\, \FUN i \vartheta (\MAT G_{10} + \MAT G_{20} + \MAT G_{30}) \big)\,  \exp\!\big(\, (\pi/4)\, \MAT G_{33} \big)\, \cdots \\
\exp\!\big(\, \vartheta (\MAT G_{01} + \MAT G_{02} + \MAT G_{03}) \big)\, \exp\!\big(\! -\vartheta (\MAT G_{10} + \MAT G_{20} + \MAT G_{30}) \big) , \M \label{eq:cartanQQ}
\end{multline}
and is therefore just a single rotation in the plane of the bivector $\MAT G_{21}$.
The action of $\MAT Q$ \& $\MAT Q'$ on the full set of $\GRP{su}(4)$ generators is summarized in Table \ref{tab:geewiz}, which indicates by their position in the table to which basis bivector each basis bivector is mapped by this transformation.

\begin{table}[htdp]
\vspace{-1ex}
\caption{ \label{tab:geewiz} \textbf{(a) left \& (b) right}}
\begin{center}
\vspace{-1ex}
\begin{tabular}{|c|c|c|c|c|} \hline
$\VEC Q^{\X[2]\dag}\VEC G_{ij}\VEC Q$ & $\cdot\,0$ & $\cdot\,1$ & $\cdot\,2$ & $\cdot\,3$ \\[1ex]\hline
$0^{\X[2]}\,\cdot$ & $\VEC G_{00}$ & $\VEC G_{23}$ & $-\VEC G_{01}$ & $-\VEC G_{22}$ \\[1ex]\hline
$1^{\X[2]}\,\cdot$ & $\VEC G_{31}$ & $-\VEC G_{12}$ & $-\VEC G_{30}$ & $-\VEC G_{13}$ \\[1ex]\hline
$2^{\X[2]}\,\cdot$ & $-\VEC G_{20}$ & $-\VEC G_{03}$ & $\VEC G_{21}$ & $\VEC G_{02}$ \\[1ex]\hline
$3^{\X[2]}\,\cdot$ & $\VEC G_{11}$ & $\VEC G_{32}$ & $-\VEC G_{10}$ & $\VEC G_{33}$ \\[1ex]\hline
\end{tabular} \M
\begin{tabular}{|c|c|c|c|c|} \hline
$\VEC Q'^{\X[2]\dag}\VEC G_{ij}\VEC Q'$ & $\cdot\,0$ & $\cdot\,1$ & $\cdot\,2$ & $\cdot\,3$ \\[1ex]\hline
$0^{\X[2]}\,\cdot$ & $\VEC G_{00}$ & $\VEC G_{01}$ & $-\VEC G_{23}$ & $~\VEC G_{22}~$ \\[1ex]\hline
$1^{\X[2]}\,\cdot$ & $\VEC G_{31}$ & $\VEC G_{30}$ & $\VEC G_{12}$ & $~\VEC G_{13}~$ \\[1ex]\hline
$2^{\X[2]}\,\cdot$ & $\VEC G_{20}$ & $\VEC G_{21}$ & $-\VEC G_{03}$ & $~\VEC G_{02}~$ \\[1ex]\hline
$3^{\X[2]}\,\cdot$ & $-\VEC G_{11}$ & $-\VEC G_{10}$ & $\VEC G_{32}$ & $~\VEC G_{33}~$ \\[1ex]\hline
\end{tabular}
\end{center}
\end{table}

Because our choice of basis in each of the two ``local'' subspaces is arbitrary, we see the essence of a transformation into the Bell basis really lies in the fact that it is a simple rotation by $\pi/2$ in a plane that meets both those subspaces orthogonally in a line.
There are of course other types of unitary transformations that can also produce entanglement, for example the SWAP or \textit{particle interchange operator},
\begin{equation}
\Pi ~\equiv~ \HALF \big( 1 + \VEC I\VEC G_{11} + \VEC I\VEC G_{22} + \VEC I\VEC G_{33} \big) ~=~ \exp\!\big( (\pi/4)\, (\VEC G_{11} + \VEC G_{22} + \VEC G_{33}) \big) .
\end{equation}
This is clearly a simultaneous rotation of one of the two qubits' subspaces onto the other, such that the chosen coordinate frames coincide.
Clearly the same rotation is obtained however these two frames are chosen, so long as they remain parallel with one another.
The generator of this rotation is also intimately connected to the singlet state, the density operator of which likewise contains it (Eq.~(\ref{eq:singlet})).
This can be regarded as an intuitive explanation for the singlet's invariance under simultaneous joint rotations of both qubits \cite{DoranLasen:03}.

\section{CLOSING REMARKS}
In this article have briefly surveyed some of the insights that our model of two qubits in the geometric algebra of a six-dimensional Euclidean vector space can provide.
There are certainly many more to come.
For one, the graph theoretical interpretation of the bivector algebra makes it clear that any bipartition of a set of orthonormal basis vectors will give rise to a Cartan decomposition, and that any maximal bipartite matching between the vertex sets will be a Cartan subalgebra.
For example, the bipartition $\{ \VEC e_{\X[1.4]1}, \VEC f_{\!1} \}$ \& $\{ \VEC e_{\X[1.4]2}, \VEC f_{\!2}, \VEC e_{\X[1.4]3}, \VEC f_{\!3} \}$ gives a subalgebra $\TEN g$ isomorphic to $\GRP{so}(2) \oplus \GRP{so}(4)$, which is seven dimensional.
A Cartan subalgebra $\TEN h$ is generated by $\VEC G_{30}, \VEC G_{03}$, providing another two degrees of freedom, and allowing us to generate all of $\GRP{SO}(6)$ with $7 + 2 + 7 = 16$ parameters.
Because $\GRP{SO}(6)$ is only $15$ dimensional, there is clearly an ineffective degree of freedom.
To identify it, we factorize $\GRP{SO}(4)$ using the Cartan decomposition $\TEN g'$ obtained from the further bipartition $\{ \VEC e_{\X[1.4]2}, \VEC f_{\!2} \}$ \& $\{ \VEC e_{\X[1.4]3}, \VEC f_{\!3} \}$, together with a matching Cartan subalgebra $\TEN h'$ generated by $\VEC G_{10}, \VEC G_{01}$.
Then since $\VEC G_{33} = \VEC e_{\X[1.4]3} \VEC f_{\!3}$ commutes with the generators $\VEC G_{30}, \VEC G_{03}$ in the first Cartan subalgebra $\TEN h$, the difference of their corresponding multipliers in the exponentials is clearly ineffective, and we have shown that $\GRP{SU}(4)$ can also be written as a product of exponentials in the following sets of generators in right-to-left order:
\begin{equation}
\{ \VEC G_{22}, \VEC G_{33} \} ,~ \{ \VEC G_{10}, \VEC G_{01} \} ,~ \{ \VEC G_{11}, \VEC G_{22} \} ,~ \{ \VEC G_{30}, \VEC G_{03}, \VEC G_{33} \} ,~ \{  \VEC G_{11}, \VEC G_{22} \} ,~ \{ \VEC G_{10}, \VEC G_{01} \} ,~ \{ \VEC G_{22}, \VEC G_{33} \} ,
\end{equation}
where we have combined the middle two copies of $\VEC G_{33} \in \TEN g'$ into the first Cartan subalgebra $\TEN h$ for simplicity.

There is also a close connection between Cartan and Schmidt decompositions, in that the latter may be obtained from the former simply by dropping all parameters that are ineffective when the transformation is applied to the projectors $P_3^1 P_3^2$.
It might be interesting to work out what the $15 - 6 = 9$ ineffective parameters are in the above Cartan factorization when it is used to derive a Schmidt-like decomposition in this way.
There are, of course, many others as well.

We have not yet paid much attention to the odd grade entities in $\EMB{\ALG G}(6)$, but there are reasons to believe they may provide additional notational efficiency in the description of various completely positive trace-preserving linear maps of $\EMB{\ALG G}^+(6)$ into itself.
For example, Kraus sums of the form
\begin{equation}
M_k \rho ~\equiv~ \rho + \tfrac14\, \VEC e_{\X[1.4]k}\, \hat\rho\, \VEC e_{\X[1.4]k} \M[2] \big( k = 1,\ldots,6\,;~ \hat\rho \equiv \rho - \tfrac14 \big)
\end{equation}
may be shown to lie on the boundary of the convex set of all completely positive trace-preserving linear maps.
By composing these maps together in all possible ways and taking convex combinations, we expect to obtain the full $96$-dimensional space of such maps.
It would be particularly interesting to see if certain physically important processes (such as the nuclear Overhauser effect in NMR \cite{ErnBodWok:87}), which cannot be described by a small number of Kraus operators in the matrix algebra, admit more concise description when sums involving odd entities are also utilized.

Finally, we point out that although the $\GRP{so}(3) \approx \GRP{su}(2)$ and $\GRP{so}(6) \approx \GRP{su}(4)$ isomorphisms do not extend to larger numbers of qubits, geometric algebra has been used to derive a complete and relativistically covariant theory of multi-qubit systems \cite{DorLasGul:93,SomLasDor:99,HavelDoran:02}.
Like the six-dimensional model dealt with here, these higher spin models introduce additional degrees of freedom which, although apparently physically irrelevant, may nevertheless be of some mathematical utility.
It could be a useful exercise to derive the six-dimensional model from within the framework of this more general theory, and gain more insight into the these extraneous degrees of freedom. 

%%%%%%%%%%%%%%%%%%%%%%%%%%%%%%%%%%%%%%%%%%%%%%%%%%%%%%%%%%%%%
%\appendix

%%%%%%%%%%%%%%%%%%%%%%%%%%%%%%%%%%%%%%%%%%%%%%%%%%%%%%%%%%%%%
\acknowledgments
This work was supported in part by a grant from the Cambridge-MIT Institute.
 
%%%%%%%%%%%%%%%%%%%%%%%%%%%%%%%%%%%%%%%%%%%%%%%%%%%%%%%%%%%%%
\bibliography{math,phys,csci,nmr,self}
\bibliographystyle{spiebib}

\end{document}